\documentclass[pre,twocolumn,showpacs]{revtex4}

\usepackage{graphicx}
\usepackage{amsmath}
\usepackage{mathtools}
\usepackage{amssymb}

\newcommand{\ba}{\begin{eqnarray}}
\newcommand{\ea}{\end{eqnarray}}
\newcommand{\be}{\begin{equation}}
\newcommand{\ee}{\end{equation}}

\begin{document}

\title{Introduction to statistical field-theory:  from a toy model to a one-component plasma
}

\author{Derek Frydel}
\affiliation{
School of Chemistry and Chemical Engineering and Institute of Natural Sciences, 
Shanghai Jiao Tong University, Shanghai 200240, China\\}
\affiliation{Laboratoire de Physico-Chime Theorique, ESPCI, CNRS Gulliver, 
10 Rue Vauquelin, 75005 Paris, France}

\date{\today}

\begin{abstract}
Working with a toy model whose partition function consists of a discrete summation, we introduce
the statistical field-theory methodology by transforming a partition function via a formal Gaussian 
integral relation (the Hubbard-Stratonovich transformation).  We then consider Gaussian type of
approximations, wherein correlational contributions enter as harmonic fluctuations around the 
saddle-point solution.  The work focuses on how to construct a self-consistent, 
non-perturbative approximation without recourse to a variational construction based on the 
Gibbs-Bogolyubov-Feynman inequality that is inapplicable to a complex action.  To address this problem, 
we propose a construction based on a selective satisfaction of a set of exact relations generated 
by considering a dual representation of a partition function, in its original and transformed form.  
\end{abstract}


\maketitle

\section{Introduction}

The treatment of electrostatics beyond the mean-field in recent years had been dominated
by the field-theoretical formalism \cite{Rudi88,Attard88,Duncan92,Netz00,Wang10,Hatlo13,Tony14}
based on the Hubbard-Stratonovich transformation of a partition function into functional integral 
over an auxiliary field.  The saddle-point of an effective Hamiltonian (or an action) within the 
transformed formulation corresponds to the mean-field solution (given by the Poisson-Boltzmann 
equation), while the harmonic fluctuations around the saddle-point constitute the 
random phase approximation treatment of correlations.  Given the charges' omnipresence
in soft-matter systems and the interest for electrostatics by workers of diverse backgrounds
(to whom the language and formalism of the field-theory is neither familiar nor intuitive) 
it is worthwhile and even desirable to present the field-theoretic formalism on a simplified 
model, where all the steps are transparent, and various challenges intrinsic to the field-theory 
formalism emphasized.  

In addition to the goal of clarity, the present article focuses on a problem of how 
to obtain a self-consistent set of equations for a non-perturbative approach, avoiding the standard 
variational 
construction based on the Gibbs-Bogolyubov-Feynman inequality (GBF), inapplicable to a complex 
action, as is the case for electrostatics.  The construction we propose is based on the hierarchy 
of exact relations extracted from dual representation of the partition function in different 
phase-spaces (the physical and the auxiliary phase-space).  The two fitting parameters of a 
Gaussian reference system, the saddle-point and the covariance matrix, are then chosen to 
satisfy any two relations of the hierarchy.    
It turns out that the first two relations within the hierarchy are automatically satisfied by using the 
variational construction based on the GBF inequality.  Our method, furthermore, opens a broader 
interpretation of the notion of self-consistency.  In principle, one can freely chose from the hierarchy
any two equations, leading to a different type of self-consistency and, in consequence, to a different 
approximation.  Thus for a two parameter Gaussian reference system different 
approximations are possible. 

The article is organized as follows.  
In Sec. \ref{sec2} we consider in detail a one-component toy model for which we develop all the 
relevant methodology.  In Sec. \ref{sec3} we generalize the model to a multicomponent system,
modifying appropriately each step.  Finally, in Sec. \ref{sec4}, we consider a realistic partition
function for a one-component plasma (a one-component system of charges in an external potential)
and derive relevant equations.

\section{The model}
\label{sec2}

A grand-canonical ensemble of a toy model that we choose to work with is given by a discrete 
summation,
\begin{equation}
\Xi = \sum_{N=0}^{\infty}\frac{\lambda^N}{N!}e^{-\varepsilon N^2/2}.
\label{eq:Xi}
\end{equation}
The crucial term in the summation is a pair interaction between particles whose strength is 
regulated by the dimensionless coupling constant $\varepsilon$.  Interactions render the exact 
analytical solution unavailable.  
An effective Hamiltonian (or an action) of a transformed partition function, obtained later in this section, 
has a mathematical structure similar to that for the system of charges, and the equations developed 
for the toy model will apply to the system of charges treated in the last section.  
$\lambda$ in  Eq. (\ref{eq:Xi}) represents a generalized 
fugacity and combines chemical and any external potential, as well as over-counting of 
self-interactions in the interaction term, $\lambda=e^{\beta\mu}e^{-\beta U_{\rm ext}}e^{\varepsilon/2}$.
Via the application of a formal identity of a Gaussian integral, 
\begin{equation}
e^{-\varepsilon N^2/2} = 
\frac{1}{\sqrt{2\pi\varepsilon}}\int_{-\infty}^{\infty}dx\,e^{-x^2/2\varepsilon}e^{ixN},
\end{equation}
the summation is transformed into integral,
\begin{equation}
\Xi = \frac{1}{\sqrt{2\pi\varepsilon}}\int_{-\infty}^{\infty} dx\,e^{-S(x)},
\label{eq:Xi_int}
\end{equation}
where the "action",
\be
S(x) = \frac{x^2}{2\varepsilon} - \lambda e^{ix},
\label{eq:action}
\ee
is a complex function, $S=S_R+iS_I$, with $S_R = x^2/2\varepsilon - \lambda\cos x,$ and 
$S_I = -\lambda\sin x$, and the resulting Boltzmann factor is a complex quantity, 
$e^{-S}=e^{-S_R}\cos S_I + ie^{-S_R}\sin S_I$, and as such eludes interpretation of a probability 
measure.  Instead a sort of "exotic" probability emerges with imaginary and negative values.  
The failure of a Boltzmann factor to fulfill the criteria of probability measure renders the application 
of a Monte Carlo sampling, used for generating various expectation values \cite{Parisi83,Klauder83}, 
no longer feasible.  In literature this is better known as the "sign" problem. 

Because the imaginary counterpart of the Boltzmann factor does not contribute to the value of a 
partition function, which is real, we write
\begin{equation}
\Xi = \frac{1}{\sqrt{2\pi\varepsilon}}\int_{-\infty}^{\infty} dx\,e^{-S_R(x)}\cos S_I(x).
\label{eq:Xi_int_R}
\end{equation}
In principle, all quantities can be obtained directly from a partition function, without the need
of imaginary functions.  On the other hand, quantities obtained as expectation values may require 
the imaginary part of the Boltzmann factor.   Let's take as an example 
\be 
-\langle ix\rangle = \frac{\int_{-\infty}^{\infty}dx\,xe^{-S_R(x)}\sin S_I(x)}{\Xi}.
\ee  
More generally, any physical expectation value admits a general form:  
\be
A(x) = A_{\rm even}(x) - iA_{\rm odd}(x).  
\ee

As a complex integral does not depend on an integration path (contour), and its value depends
on endpoints only, we are free to select an integration path $C$, 
\begin{equation}
\Xi=\frac{1}{\sqrt{2\pi\varepsilon}}\int_C dz\,e^{-S(z)},
\end{equation}
in which case $S$ becomes a function of a complex variable, $z=x+iy$, 
\be
S_R(x,y)=\frac{x^2-y^2}{2\varepsilon} - \lambda e^{-y}\cos x,
\ee
and
\be
S_I(x,y) = \frac{xy}{\varepsilon} - \lambda e^{-y}\sin x.  
\label{eq:S_I}
\ee

\subsection{the saddle-point approximation}
\begin{figure}[h] 
 \begin{center}%
 \begin{tabular}{rr}
  \includegraphics[height=0.3\textwidth,width=0.45\textwidth]{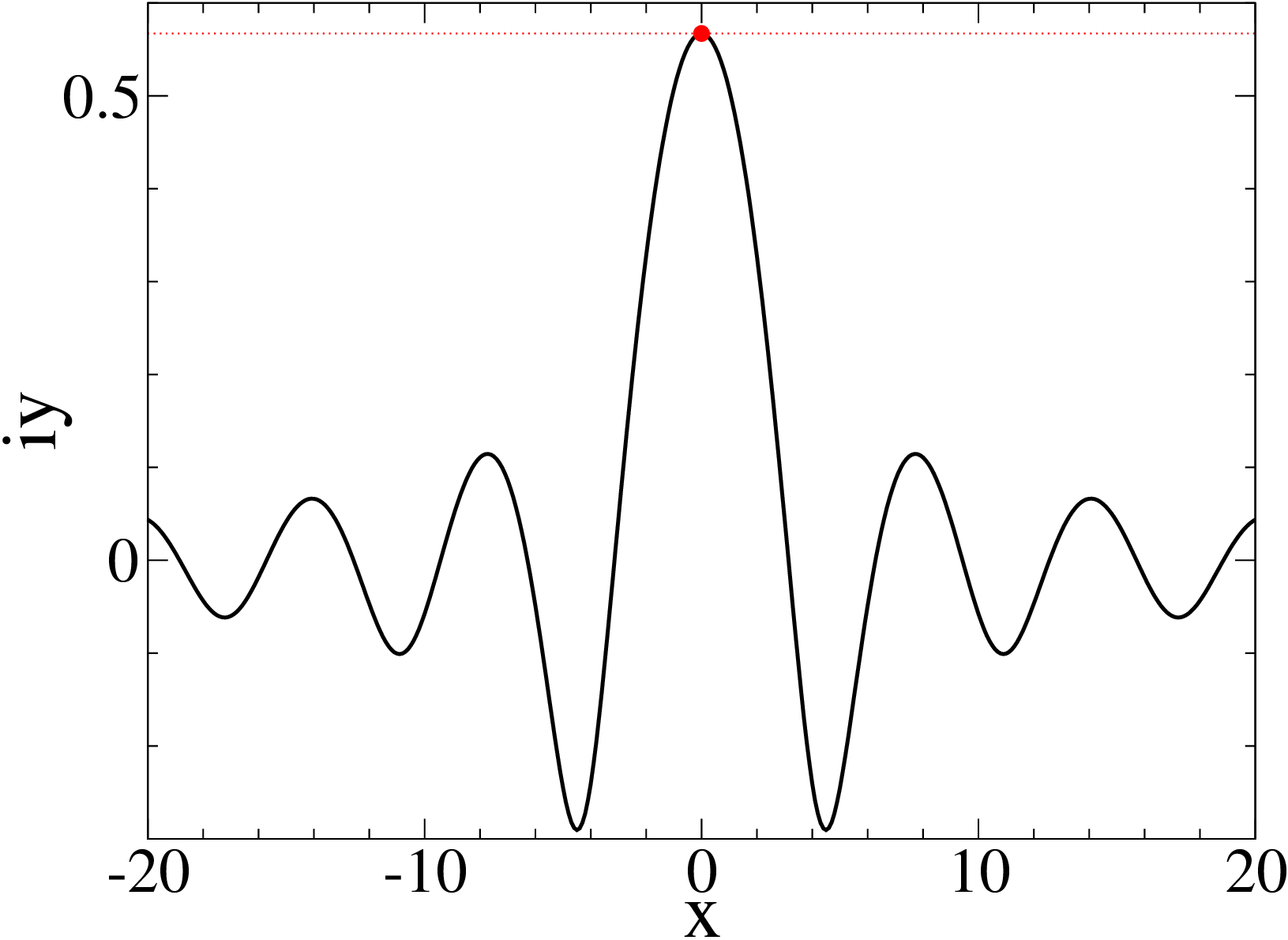}\\
\end{tabular}
 \end{center}
\caption{The integration path $C$ on the complex plane, where $S_I(z)=0$ for all $z\in C$.  
$C$ corresponds to the solution of Eq. (\ref{eq:y_x}) for $\varepsilon\lambda=1$.
In addition we plot a contour $z\in [-\infty+iy_0,\infty+iy_0]$ parallel to the real axis and 
passing through the saddle-point $z_0$.}
\label{fig:C}
\end{figure}

A preferred integration contour should have small or no oscillations of a Boltzmann factor.  
Along such a contour the imaginary part (or phase) of an action should be suppressed, $S_I(x)\to 0$.
If a contour satisfies 
\be
S_I(z) = 0,
\ee
for every $z \in C$, the partition function can be written as
\be
\Xi=\int_{C} dz\,e^{-S_R}.
\ee
From Eq. (\ref{eq:S_I}), the constraint $S_I(z)=0$ leads to the equation
\be
y = \varepsilon\lambda e^{-y}\bigg(\frac{\sin x}{x}\bigg),
\label{eq:y_x}
\ee
where the solution is
\be
y(x) = W\bigg(\frac{\varepsilon\lambda\sin x}{x}\bigg),
\label{eq:C}
\ee
and where $W(x)$ is the Lambert function.  We plot $W(\varepsilon\lambda\sin x/x)$ for 
$\varepsilon\lambda=1$ in Fig.(\ref{fig:C}).  Oscillations quenched in the Boltzmann factor come 
back in the contour.  
\graphicspath{{figures/}}

\begin{figure}[h] 
 \begin{center}%
 \begin{tabular}{rr}
  \includegraphics[height=0.3\textwidth,width=0.45\textwidth]{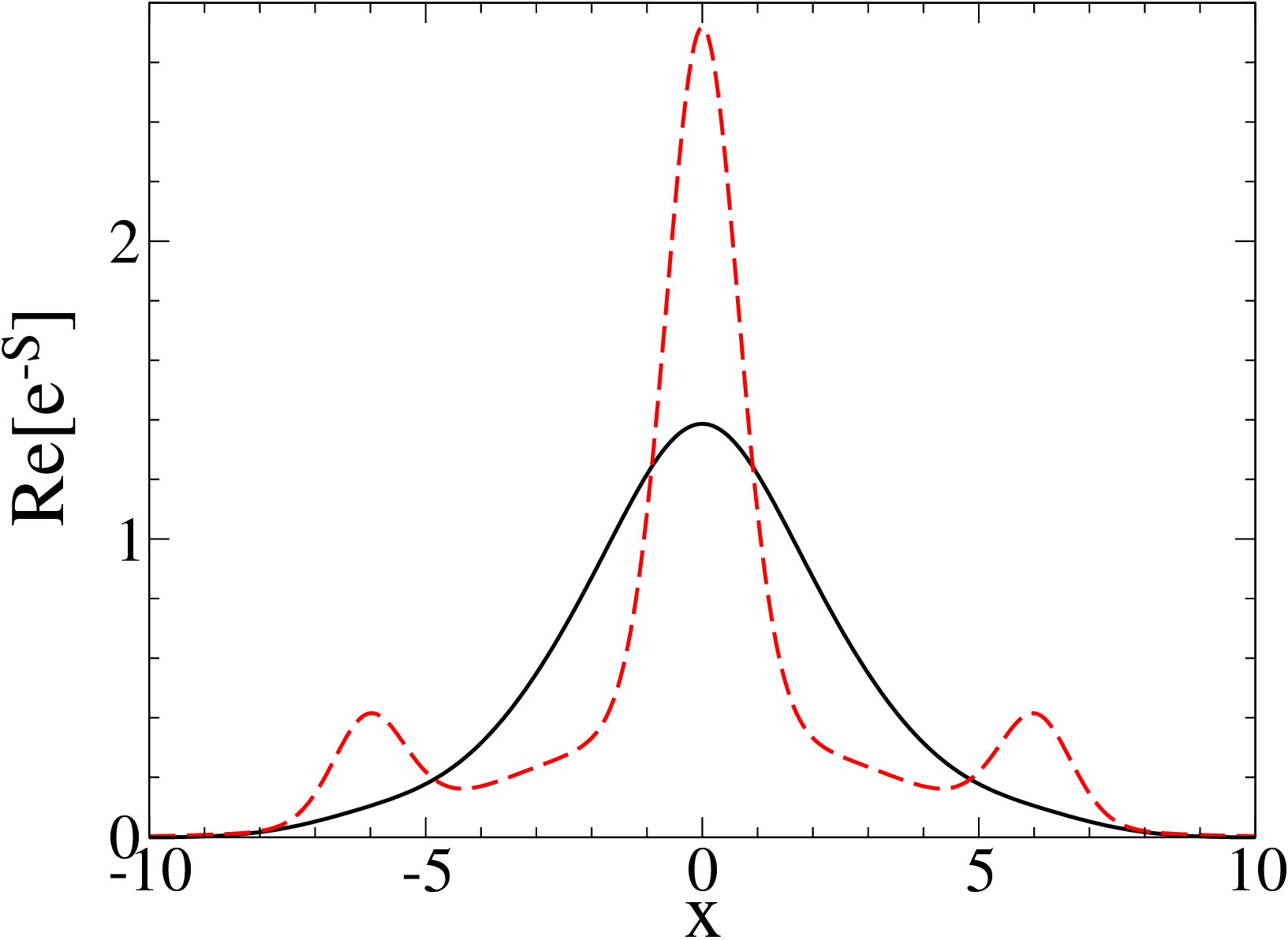}\\
\end{tabular}
 \end{center}
\caption{The real part of the Boltzmann factor, ${\rm Re}[e^{-S(z)}]$, for two different contours:  
$z\in[-\infty,\infty]$
(dashed line) and $z\in[-\infty+y_0,\infty+y_0]$ (solid line).  
The system parameters are $\varepsilon=1$ and $\lambda=10$.}
\label{fig:P_R}
\end{figure}

A simpler path, with reduced oscillations in the action, 
is a path parallel to the real axis but displaced along the imaginary axis to include the 
saddle-point,
\be
\frac{\partial S}{\partial z}\bigg|_{z_0} = 0.  
\ee
The condition of stationarity yields
\be
y_0 = \lambda\varepsilon e^{-y_0}.
\label{eq:y_0}
\ee
In Fig. (\ref{fig:P_R}) we plot ${\rm Re}\,[e^{-S}]$ for contours parallel to the real axis
and displaced along the imaginary axis by $y=0$ and $y=y_0$.  The contour that includes
the saddle-point displays reduced oscillations.    
\graphicspath{{figures/}}


A physical interpretation of the saddle-point is gained by considering the expectation value 
$\langle iz\rangle$,
\ba
-\langle iz\rangle &=& y_0-\langle ix\rangle_{y_0}\nonumber\\ 
&=& y_0 + \frac{1}{\Xi}\int_{-\infty}^{\infty} dx\,x e^{-S_R(x,y)}\sin S_I(x,y).\nonumber\\
\label{eq:iz}
\ea
The approximation $-\langle iz\rangle\approx y_0$, given by the saddle-point approximation,  
becomes more accurate if the second term is negligible.  The saddle-point solution is identified with 
the mean-field approximation where the action is rewritten in terms of average quantities, 
\be
S_{\rm mf} = -\frac{\langle ix\rangle_{\rm mf}^2}{2\varepsilon} - \lambda e^{\langle ix\rangle_{\rm mf}},
\ee
so that $\Xi_{\rm mf}\sim e^{-S_{\rm mf}}$ and the quantity $\langle ix\rangle_{\rm mf}$ ought to
minimize $\Xi_{\rm mf}$, which yields
\be
-\langle ix\rangle_{\rm mf} = \lambda e^{\langle ix\rangle_{\rm mf}},
\ee
and recovers the relation in Eq. (\ref{eq:y_0}).

\subsection{Gaussian fluctuations}
\label{sec:perturbative_gauss}
Within a Gaussian approximation a partition function is approximated by expanding $S$ around the 
saddle-point up to a harmonic term,
\be
S(z) \approx S(z_0) + \frac{1}{2}S''(z_0)(z-z_0)^2.
\label{eq:f_gauss}
\ee
This generates a Gaussian functional form of the Boltzmann factor,
\be
e^{-S(z)}=e^{-S(z_0)}e^{-(z-z_0)^2/(2\Gamma)}
\ee
with variance
\be
\Gamma = \frac{1}{S''(z_0)} = \frac{\varepsilon}{1+\varepsilon\lambda e^{-y_0}}.
\label{eq:variance}
\ee
For the contour 
intercepting the saddle-point $y_0$, the approximation in Eq. (\ref{eq:f_gauss}) becomes
\be
S(z) \approx S_R(z_0) + \frac{x^2}{2\Gamma}, 
\ee
and the integrand becomes $e^{-x^2/2\Gamma}$, without 
oscillations, and the approximate partition function is
\ba
\Xi_{\rm G} &=& e^{-S(z_0)}\int_{-\infty}^{\infty} \frac{dx}{\sqrt{2\pi\varepsilon}}\,e^{-x^2/2\Gamma}\nonumber\\
&=& e^{-S(z_0)}\sqrt{\frac{\Gamma}{\varepsilon}}.  
\label{eq:Xi_sp}
\ea
The corresponding grand potential is
\begin{equation}
\beta\Omega_{\rm G} = S(z_0) - \log\sqrt{\frac{\Gamma}{\varepsilon}}.
\end{equation}
Within this approximation $-\langle iz\rangle=y_0$, as for the mean-field.  (The second term in 
Eq. (\ref{eq:iz}) is zero since for the Gaussian approximation the imaginary part is suppressed).   
The fluctuations, on the other hand, are given by a variance, $\langle \delta z^2\rangle=\Gamma$.

A more accurate approximation is possible if one takes an alternative definition for 
$\langle iz\rangle$, 
\be
-\langle iz\rangle=\varepsilon\lambda\frac{\partial\beta\Omega_G}{\partial\lambda},
\ee
which yields
\ba
-\langle iz\rangle
&=&y_0-\frac{\varepsilon\lambda}{2}\frac{\partial\log\Gamma}{\partial\lambda}\nonumber\\
&=&y_0 - \frac{1}{2}\lambda e^{-y_0}\Gamma^2,
\label{eq:iz_3}
\ea
where the harmonic fluctuations reduce the value of the mean-field.

\subsection{higher-order fluctuations}
One wonders how higher-order fluctuations, beyond the harmonic term, contribute to the 
approximation.  Accordingly, we write
\be
S_3(z) = S(z_0) + \frac{(z-z_0)^2}{2\Gamma} + \frac{S'''(z_0)}{6}(z-z_0)^3,
\ee
where $S'''(z_0)=i\lambda e^{-y_0}$.  
For the contour along the real axis and displaced by $y_0$ we get
\be
S_3(z) = S_R(z_0) + \frac{x^2}{2\Gamma} + i\bigg(\frac{\lambda e^{-y_0}}{6}\bigg)x^3,
\ee
and the resulting Boltzmann factor has now a phase and associated with it oscillations, due
to the imaginary term.  The resulting partition function is
\ba
\Xi_{\rm 3} 
&=& e^{-S(z_0)}\int_{-\infty}^{\infty} \frac{dx}{\sqrt{2\pi\varepsilon}}\,
e^{-x^2/2\Gamma}e^{-i\lambda e^{-y_0}x^3/6}.
\ea
If we simplify the expression, using $e^{a}\approx 1+a$, we get
\be
\Xi_3 \approx e^{-S(z_0)}\int_{-\infty}^{\infty} \frac{dx}{\sqrt{2\pi\varepsilon}}\,
e^{-x^2/2\Gamma}\bigg(1-\frac{i}{6}\lambda e^{-y_0}x^3\bigg).
\ee
and the expectation value $\langle iz\rangle$ is
\ba
-\langle iz\rangle &=& y_0 - \frac{1}{6}\frac{\lambda e^{-y_0}}{\Xi_G}
\int_{-\infty}^{\infty} \frac{dx}{\sqrt{2\pi\varepsilon}}\,x^4 e^{-x^2/2\Gamma}\nonumber\\
&=&y_0 - \frac{1}{2}\lambda e^{-y_0}\Gamma^2.  
\ea
The result is exactly the same as that in Eq. (\ref{eq:iz_3}).  

\begin{figure}[h] 
 \begin{center}%
 \begin{tabular}{rr}
  \includegraphics[height=0.25\textwidth,width=0.4\textwidth]{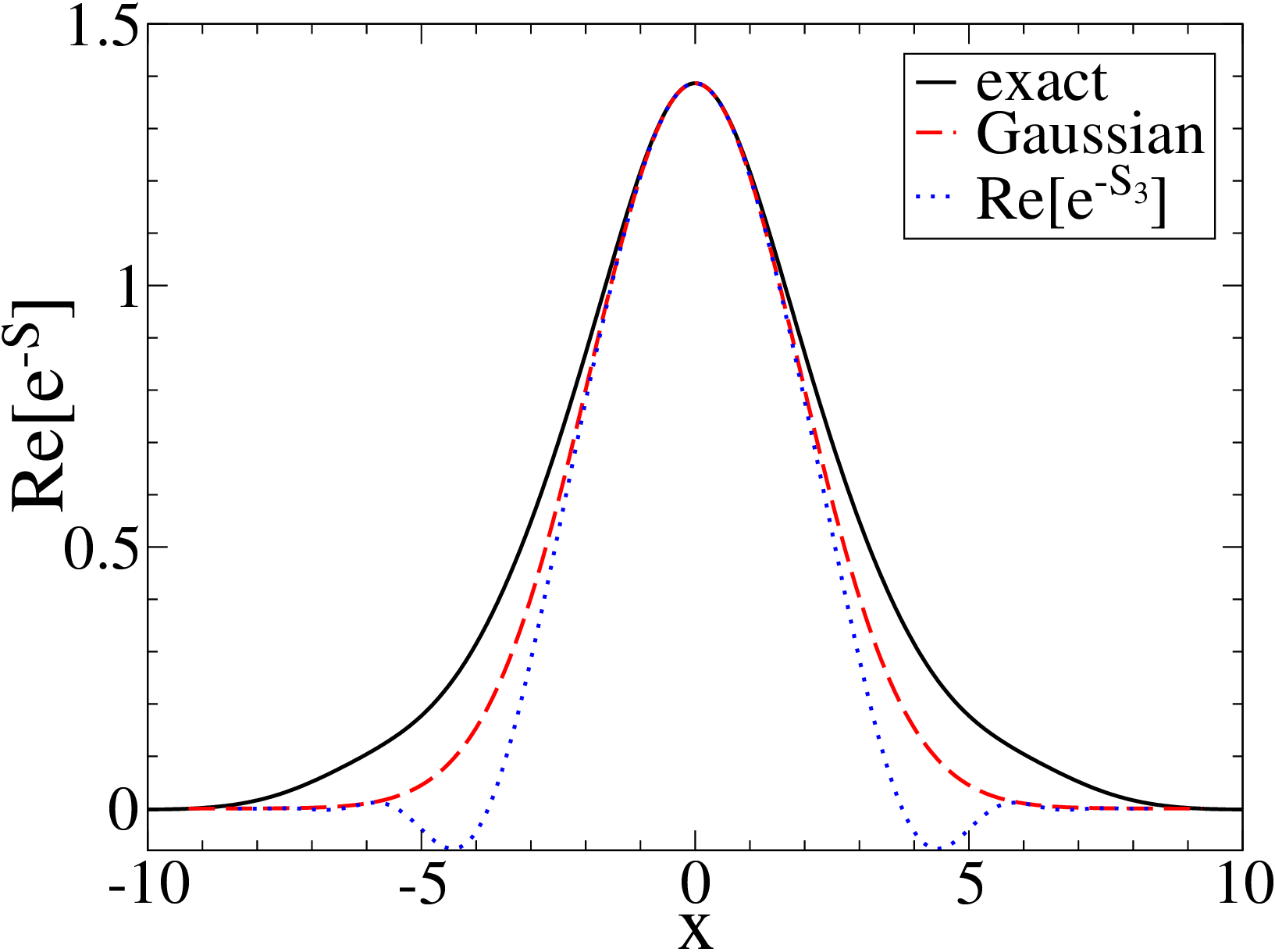}\\
\end{tabular}
 \end{center}
\caption{A Boltzmann factor, ${\rm Re}[e^{-S}]$, plotted along the real axis displaced by
$y_0$ along the imaginary direction. 
The system parameters are $\lambda=1$ and $\varepsilon=10$. }
\label{fig:P3}
\end{figure}

However, using a complete expression $\Xi_3$ we find worsening of the results.  
In Fig. (\ref{fig:P3}) we plot various fluctuation distributions around the saddle-point.  
It is clear from the figure that the inclusion of the next term beyond the harmonic term 
gives rise to unphysical oscillations that deviate from the exact curve.  
\graphicspath{{figures/}}

\subsection{non-perturbative approach}
In an alternative procedure, a Gaussian approximation is improved non-perturbatively,
that is, an action is still expanded up to a harmonic term, 
\be
S_g(z) \approx  \frac{(z-z_{0})^2}{2\Gamma_{}},
\ee
but the parameters $z_0$ and $\Gamma$ play a role of a fitting parameter.  The standard construction 
to obtain these parameters is based on a variational procedure based on the Gibbs-Bogolyubov-Feynman 
inequality (GBF), 
\be
\Xi = \Xi_g\langle e^{-\Delta S}\rangle_g\ge \Xi_g e^{-\langle \Delta S\rangle_g},
\label{eq:Xi_GBF}
\ee
where
\be
\langle\dots\rangle_g = \frac{\int dx\,e^{-S_g(x)}[\dots]}{\int dx\,e^{-S_g(x)}},
\ee
and 
\be
\Xi_g = \int dx\,e^{-S_g(x)}, 
\ee
and $\Delta S = S-S_g$.  The variational procedure, thus, provides an upper bound for a free 
energy and the sought parameters are obtained from minimization.  
The inequality in Eq. (\ref{eq:Xi_GBF}) is obtained from a more basic
inequality, $e^{a}\ge 1+a$, where $a$ is a real number.  If $\Delta S$ were real, we could write 
$$
\langle e^{-\Delta S}\rangle = e^{-\langle\Delta S\rangle}
\big\langle e^{-(\Delta S-\langle \Delta S\rangle)}\big\rangle
\ge e^{-\langle\Delta S\rangle},
$$
However, the problem is that $\Delta S$ of the toy model 
is complex and the above inequality does not hold.  
One consequence is that the minimization principle is replaced by the stationarity principle 
\cite{Spruch83,Rau06,Ruel08}, where the true value is not approached from above but in oscillatory
fashion, either from above and below, violating the GBF inequality.  

To get around this difficulty we propose a different approach, where self-consistency is enforced
by making sure that some known identities are satisfied.  To derive these identities for the present
toy model, we recall two alternative formulations of the partition function:
\ba
\Xi &=& 
\sum_{N=0}^\infty\frac{\lambda^N}{N!}e^{-\frac{\varepsilon}{2}(N-\varrho)^2}\nonumber\\
&=&\int_{-\infty}^{\infty}\frac{dx}{\sqrt{2\pi\varepsilon}}dx\,e^{-S(x)-i\varrho x},
\ea
where we have introduced a source term, $S\to S+i\varrho x$, which in a final expression is
taken to zero.  From the original formulation of the partition function we know 
\be
\lambda\frac{\partial\log\Xi}{\partial\lambda} = \frac{1}{\varepsilon}\frac{\partial\log\Xi}{\partial\varrho} + \varrho.
\label{eq:identity_1}
\ee
When applied to the field-theoretical formulation this relation yields (in the limit $\varrho\to 0$)
\be
\varepsilon\lambda\langle e^{iz}\rangle = -\langle iz\rangle.
\label{eq:ident_1}
\ee
One can generate an arbitrary number of additional identities by repeated application of derivatives 
$\frac{\partial}{\partial\rho}$ and $\frac{\partial}{\partial\lambda}$ to both sides of Eq. (\ref{eq:identity_1}).  
The first three identities are
  \ba
 &&-\big\langle iz\big\rangle = \varepsilon\lambda \big\langle e^{iz}\big\rangle\nonumber\\
 &&-\varepsilon\lambda\big\langle(\delta e^{iz})(i\delta z)\big\rangle = \varepsilon - \big\langle \delta z^2\big\rangle
 \nonumber\\
 &&-\varepsilon\lambda\big\langle(\delta e^{iz})(i\delta z)\big\rangle = 
 \varepsilon^2\lambda\big\langle e^{iz}\big\rangle
 +\varepsilon^2\lambda^2 \big\langle (\delta e^{iz})^2\big\rangle
 \nonumber\\&&~~~~~~~~~~~~~~~~~~~~\dots
\label{eq:relations}
  \ea   
where $\delta z=z-\langle z\rangle$ and $\delta e^{iz}=e^{iz}-\langle e^{iz}\rangle$.  The second and the 
third equation provide a relation between different second-order fluctuations.  By continually applying the 
derivatives, one obtains equations for the third- and higher-order fluctuations.  We are not interested 
in these higher-order relations.  

Within the mean-field approximation the first identity yields
\be
y_0 = \varepsilon\lambda e^{-y_0},
\ee
where the distribution over the auxiliary phase-space is a delta function at the saddle-point.
The absence of fluctuations trivially satisfies the second and third equation as $0=0$.  
Moving on to a Gaussian distribution, we start by listing relevant expectation values 
\ba
&&\langle iz\rangle_g = -y_0\nonumber\\
&&\langle e^{i z}\rangle_g = e^{-y_0}e^{-\Gamma/2}\nonumber\\
&&\langle \delta z^2\rangle_g = \Gamma\nonumber\\
&&\langle (\delta iz)(\delta e^{iz})\rangle_g = -\Gamma e^{-y_0}e^{-\Gamma/2}\nonumber\\
&&\langle (\delta e^{iz})^2\rangle_g = e^{-2y_0}e^{-\Gamma}(e^{-\Gamma}-1),\nonumber\\
\label{eq:expectations_gauss}
\ea
and the three equalities in Eq. (\ref{eq:relations}) become
 \ba
 &&y_0 = \varepsilon\lambda e^{-y_0}e^{-\Gamma/2},\nonumber\\
 &&\Gamma\big(\varepsilon\lambda e^{-y_0}e^{-\Gamma/2}\big) = \varepsilon-\Gamma \nonumber\\
 &&(1-e^{-\Gamma})\big(\varepsilon\lambda e^{-y_0}e^{-\Gamma/2}\big) = \varepsilon - \Gamma
 \nonumber\\&&~~~~~~~~~~~~~~~~~\dots
 \label{eq:relations_gauss}
  \ea   
The perturbative approach of Sec. (\ref{sec:perturbative_gauss}) does not produce self-consistent
relations.  Self-consistency has to be enforced "by hand" by choosing appropriate values for 
$y_0$ and $\Gamma$.  For a two parameter model, one equation in Eq. (\ref{eq:relations_gauss})
is superfluous.  Closer examination reveals that the second equation is a linear version of the third.  

Incidentally, the variational construction based on the GBF automatically satisfies the first two 
equations.  By making this identification, we provide alternative justification and provide different 
interpretation to the GBF variational construction.  Moreover, with other identities available, 
one can choose different set of identities that yield a different self-consistent non-variational 
scheme.

\begin{figure}[h] 
 \begin{center}%
 \begin{tabular}{rr}
  \includegraphics[height=0.25\textwidth,width=0.4\textwidth]{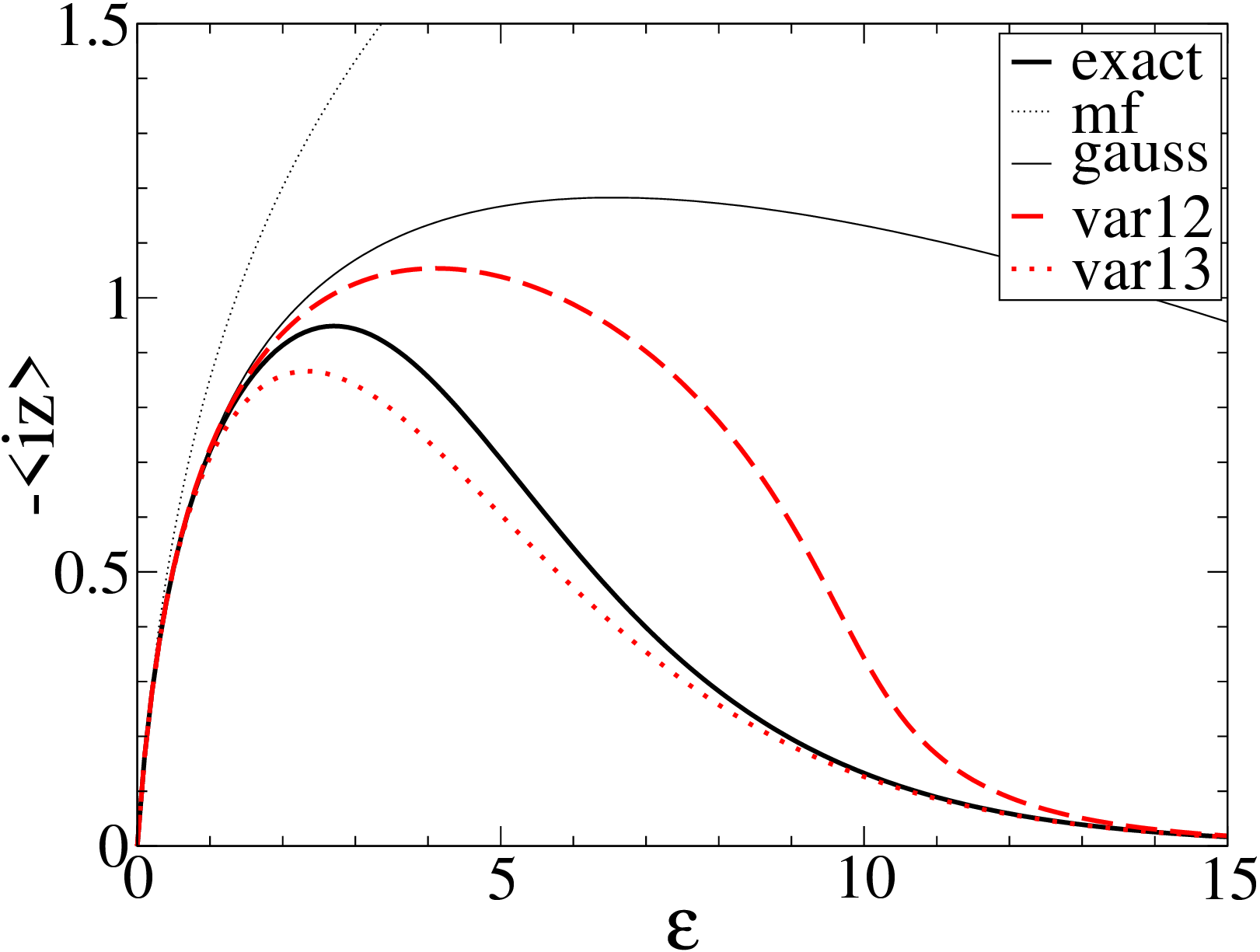}\\
  \includegraphics[height=0.25\textwidth,width=0.4\textwidth]{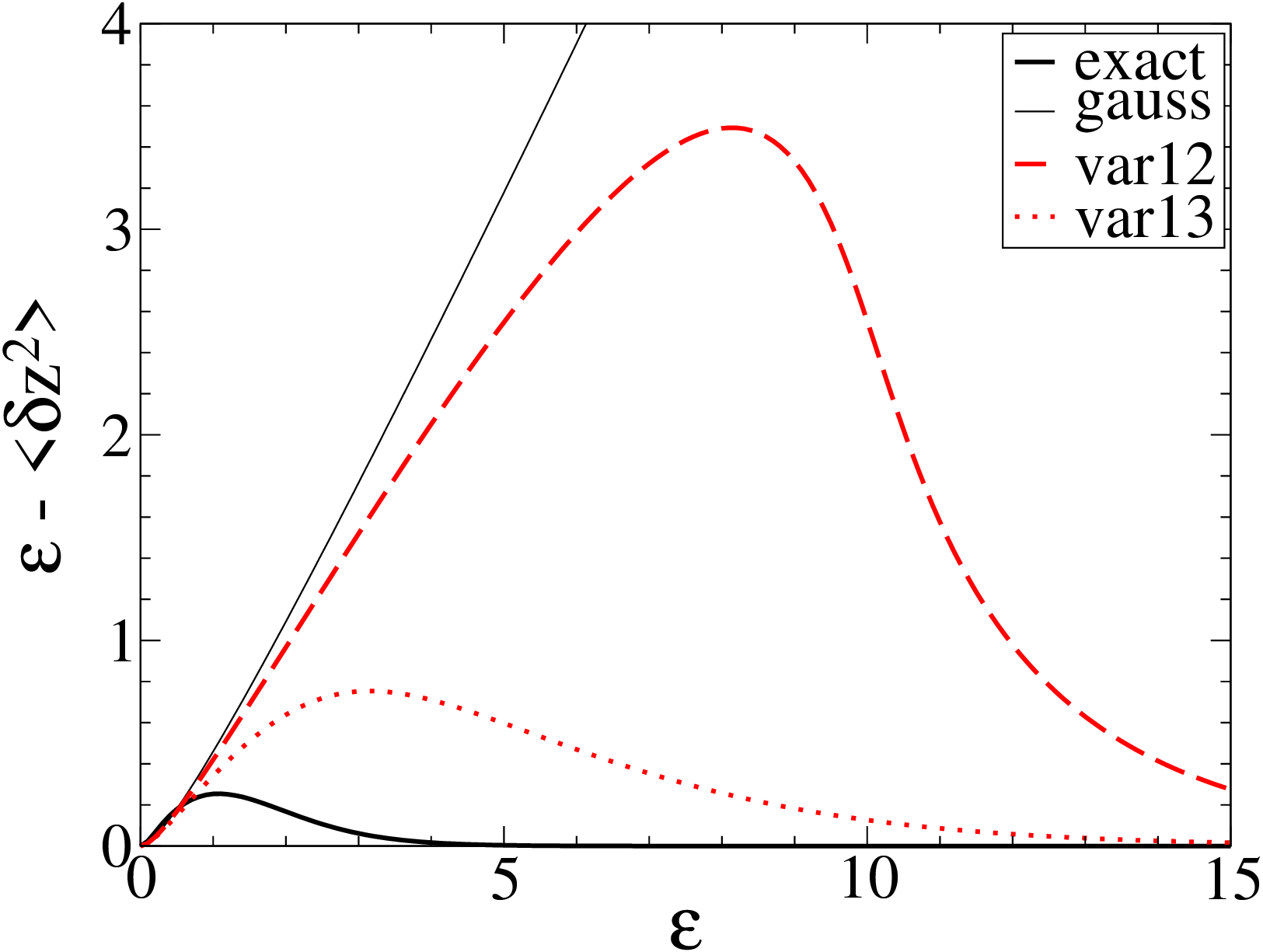}\\
\end{tabular}
 \end{center}
\caption{The expectation values -$\langle iz\rangle$ and $\langle \delta z^2\rangle$ as a function of the
coupling constant $\varepsilon$ for $\lambda=2$.  These expectation values give an average number
of particle and their fluctuations:
$-\langle iz\rangle=\varepsilon\langle N\rangle$
and $\varepsilon-\langle\delta z^2\rangle = \varepsilon^2\langle \delta N^2\rangle$.}
\label{fig:y_L2}
\end{figure}

In Fig. (\ref{fig:y_L2}) we compare different schemes by plotting various expectation values.  
Self-consistent schemes are more accurate for a large coupling constant.  
Furthermore, the approach based on equations $(1,3)$ (line 1 an 3 in Eq. (\ref{eq:relations_gauss})) 
is more accurate than that based on 
equations $(1,2)$ (that is equivalent to the GBF variational construction).  
The model and the test, however, are too simple to generalize the conclusion to all 
cases.  In the subsequent section we consider a multiple species model, as a way to introduce
complexity.  However, we first discuss some aspects of the present simple model.  
\graphicspath{{figures/}}

\begin{figure}[h] 
 \begin{center}%
 \begin{tabular}{rr}
  \includegraphics[height=0.25\textwidth,width=0.4\textwidth]{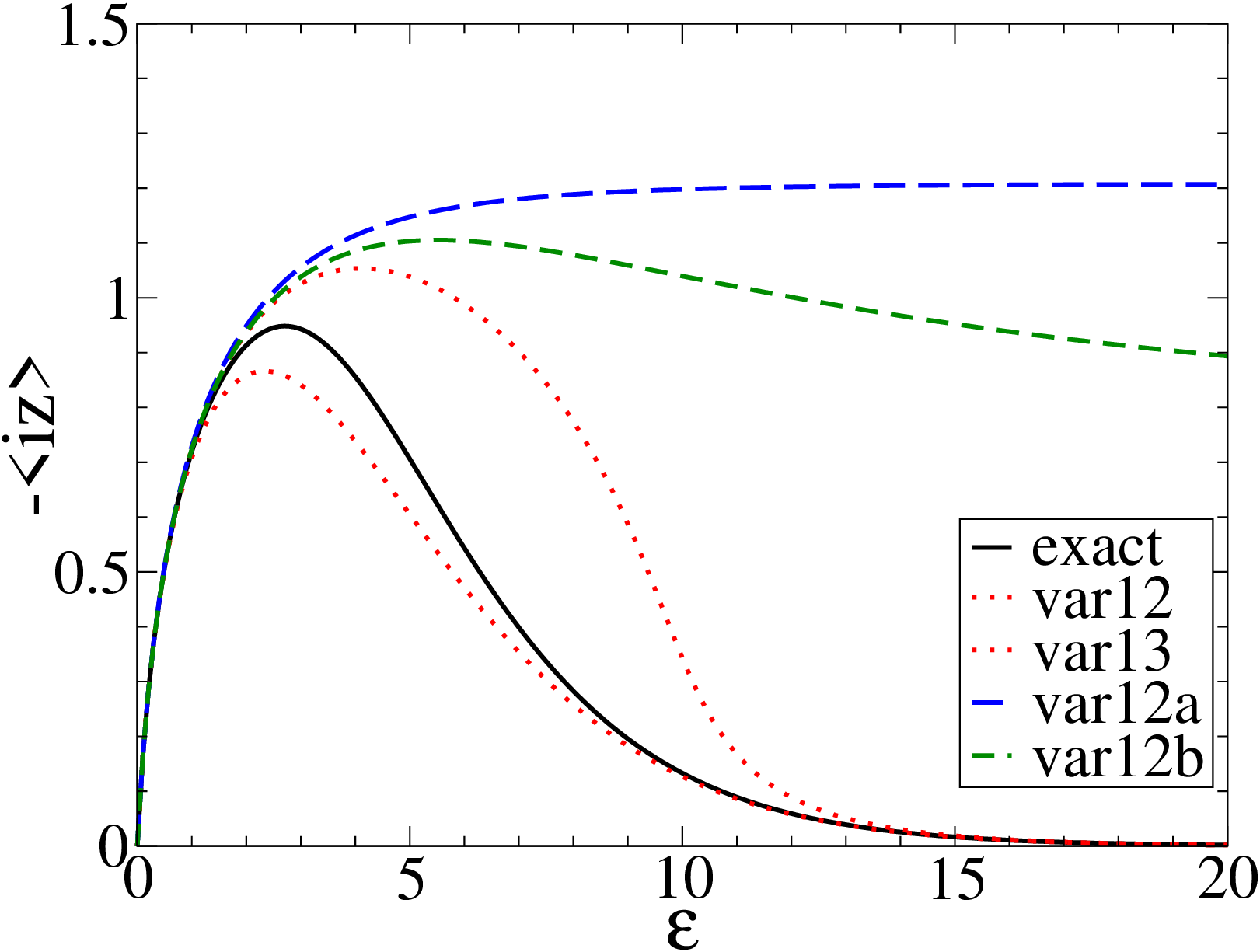}\\
\end{tabular}
 \end{center}
\caption{The expectation values -$\langle iz\rangle$  as a function of the
coupling constant $\varepsilon$ for $\lambda=2$, for exponential ("var12a") and square ("var12b")
distributions.  }
\label{fig:y_L2a}
\end{figure}

One possible question is:  is a Gaussian distribution the best representation of a true distribution.  Practical 
concerns demand that a distribution affords analytical solutions.  A Gaussian distribution satisfies
this requirement, and so it is a convenient tool.  But the simplicity of the present model allows us to 
explore other distributions, such as a stretched Gaussian distribution:  
$e^{-[(z-z_0)^2/2\Gamma]^{\gamma}}$, where $\gamma=0.5$ corresponds to an exponential, 
$\gamma\to\infty$ to a square, and $\gamma=1$ to a Gaussian distribution.  Below we provide relevant 
expectation values for an exponential distribution (compare with Eq. (\ref{eq:expectations_gauss}) for the 
Gaussian case): 
\ba
&&\langle iz\rangle_{exp} = -y_0\nonumber\\
&&\langle e^{i z}\rangle_{exp} = \frac{e^{-y_0}}{1+2\Gamma}\nonumber\\
&&\langle \delta z^2\rangle_{exp} = 4 \Gamma\nonumber\\
&&\langle (\delta iz)(\delta e^{iz})\rangle_{exp} = -\frac{4\Gamma e^{-y_0}}{(1+2\Gamma)^2}\nonumber\\
&&\langle (\delta e^{iz})^2\rangle_{exp} = \frac{e^{-2y_0}}{1+8\Gamma}-\frac{e^{-2y_0}}{(1+2\Gamma)^2}.\nonumber\\
\ea
Plugging these expressions into corresponding equations in Eq. (\ref{eq:relations}), we may 
obtain $y_0$ and $\Gamma$.  The results for the exponential and square distributions are shown in 
Fig. (\ref{fig:y_L2a}) (labeled as "var12a" and "var12b", respectively).  The new curves are less accurate 
than those for a Gaussian distribution ("var12" or "var13").  On the other hand, their results are not inferior 
to a perturbative Gaussian approach.  This seems to show that self-consistency is an important element 
of an approximation.  
\graphicspath{{figures/}}

Another question is, isn't a three parameter model more natural, as there are three available equations.  
Even if disregarding complexity, we found no improvement 
for a number of different three parameter models.  This seems to imply that a three parameter model 
requires the identity involving fluctuations of a third-order,
\ba
&&~~~~~~~~~~~~~~~~~~~~~~~~~~~~~\dots\nonumber\\
&&\lambda\big\langle (i\delta z)^2(\delta e^{iz})\big\rangle=-\frac{1}{\varepsilon}\big\langle (i\delta z)^3\big\rangle\nonumber\\
&&-\big\langle (i\delta z)(\delta e^{iz})\big\rangle - \lambda\big\langle (i\delta z)(\delta e^{iz})^2\big\rangle
=\frac{1}{\varepsilon}\big\langle(i\delta z)^2(\delta e^{iz})\big\rangle\nonumber\\
&&2\big\langle(\delta e^{iz})^2\big\rangle + \lambda\big\langle (\delta e^{iz})^3\big\rangle
=-\frac{1}{\varepsilon}\big\langle (i\delta z)(\delta e^{iz})^2\big\rangle\nonumber\\
&&~~~~~~~~~~~~~~~~~~~~~~~~~~~~~\dots
\ea
This, however, would markedly complicate the calculations, and we stop here.

\section{Multiple species}
\label{sec3}

In this section we consider a partition function for multiple species,
\ba
\Xi &=& 
\sum_{\{N_i\}}
\prod_{i=1}^{M}\bigg(\frac{\lambda_{i}^{N_i}}{N_i!}\bigg)e^{-\frac{1}{2}\sum_{i,j}{\mathcal E}_{ij}N_iN_j}
\nonumber\\&=&
\sum_{\{N_i\}}\prod_{i=1}^{M}\bigg(\frac{\lambda_{i}^{N_i}}{N_i!}\bigg)\,
e^{-\frac{1}{2}({\bf N}^{T}{\mathcal E}{\bf N})},
\label{eq:Xi_N}
\ea
where $M$ is the number of different species, and interactions between species are regulated by  
elements of the $M\times M$ connectivity matrix, ${\mathcal E}_{ij}$.  The number of particles of each
species span the values $N_i=0,1,2,\dots,\infty$.  

Using a formal identity of a Gaussian integral,
\begin{equation}
e^{-\frac{1}{2}({\bf N}^{T}{\mathcal E}{\bf N})} = 
\int\frac{d{\bf x}}{\sqrt{\det{2\pi\mathcal E}}}\,e^{-\frac{1}{2}({\bf x}^{T}{\mathcal E}^{-1}{\bf x})}
e^{i{\bf N}^{T}{\bf x}},
\label{eq:identity}
\end{equation}
the partition function is rewritten as
\begin{equation}
\Xi = \int \frac{d{\bf x}}{\sqrt{\det{2\pi\mathcal E}}}\,e^{-S({\bf x})},
\label{eq:Xi_FT_N}
\end{equation}
where the action is
\begin{equation}
S({\bf x}) = \frac{{\bf x}^{T}{\mathcal E}^{-1}{\bf x}}{2}
-\sum_{i=1}^M\lambda_ie^{ix_i}.
\label{eq:f_N}
\end{equation}
${\mathcal E}^{-1}$ is the inverse of ${\mathcal E}$, ${\bf x}\equiv\{x_1,x_2,\dots,x_M\}$ is a vector, and 
\begin{equation}
\int d{\bf x}=\int_{-\infty}^{\infty}\prod_{i=1}^{M}dx_i.
\end{equation}

\subsection{the saddle-point}
As for the single species mode, we generalize the real vector ${\bf x}$ to its complex counterpart, 
${\bf z}={\bf x}+i{\bf y}$.  The saddle-point now corresponds to a vector ${\bf z}_0$ at which $S$ 
is stationary,
\begin{equation}
\frac{\partial S({\bf z})}{\partial {\bf z}}\bigg|_{{\bf z}={\bf z}_0} = 0.
\label{eq:sp_M}
\end{equation}
and, as for the case $M=1$, a solution is strictly imaginary, 
${\bf z}_0=i{\bf y}_0$, where
\begin{equation}
y_{i} = \sum_{i=1}^{N}{\mathcal E}_{ij}\lambda_ie^{-y_{j}}.
\label{eq:yM_0}
\end{equation}
To simplify the nomenclature, we use $y_i$ to indicate the saddle-point value, $y_{i}\equiv y_{0,i}$.

\subsection{Gaussian fluctuations}
By expanding the action up to a harmonic term, we capture fluctuations, 
\begin{equation}
S({\bf z})\approx S({\bf z}_0) + 
\frac{1}{2}\sum_{i,j}^N\frac{\partial^2 S({\bf z}_0)}{\partial z_i\partial z_j} 
(z_i-z_{0,i})(z_j-z_{0,j}),
\end{equation}
and the partition function becomes
\begin{eqnarray}
\Xi_{\rm g} &=& \frac{e^{-S({\bf z}_0)}}{\sqrt{\det 2\pi{\mathcal E}}}
\int_C d{\bf z}\, 
e^{-\frac{1}{2\varepsilon}({\bf z}-{\bf z}_0)^T{\Gamma}^{-1}({\bf z}-{\bf z}_0)}\nonumber\\
&=&e^{-S({\bf z}_0)}\sqrt{\det{\Gamma}{\mathcal E}^{-1}},
\end{eqnarray}
where 
\begin{equation}
{\Gamma}_{ij}^{-1} = 
\frac{\partial^2 S({\bf z}_0)}{\partial z_i\partial z_j}
={\mathcal E}^{-1}_{ij} + \lambda_i e^{iz_{0,i}}\delta_{ij}
\label{eq:S_G}
\end{equation}
is a covariance matrix.  

Gaussian fluctuations correct the mean-field value for $\langle iz_i\rangle$ if we use a definition 
$-\langle iz_i\rangle=\frac{\partial\ln\Xi_{\rm}}{\partial\varrho_i}$, where the source 
is implemented into the action, $S\to S +i\boldsymbol{\varrho}\cdot{\bf z}$, yielding
\begin{eqnarray}
-\langle iz_i\rangle 
&=& \frac{\partial S({\bf z}_0)}{\partial\varrho_i} 
- \frac{\partial\ln\sqrt{\det{\Gamma}}}{\partial\varrho_i}\nonumber\\
&=& y_{i} - \sum_{j=1}^M\frac{\partial\ln\sqrt{\det{\Gamma}}}{\partial y_{j}}
\frac{\partial y_{j}}{\partial\varrho_{i}}\nonumber\\
&=&y_{i} - \frac{1}{2}
\sum_{j=1}^M\lambda_je^{-y_{j}}{\Gamma}_{jj}{\Gamma}_{ij},
\label{eq:izM_3}
\end{eqnarray}
in the limit $\boldsymbol\varrho\to {\bf 0}$.  
Compare with Eq. (\ref{eq:iz_3}) for $M=1$.

\subsection{non-perturbative approach}
We construct a non-perturbative approximation based on a Gaussian reference system,
\begin{equation}
\Xi_0 = \int _C\frac{d{\bf z}}{\sqrt{\det 2\pi\mathcal E}}
e^{-\frac{1}{2}({\bf z}-{\bf z}_0)^T{\Gamma}^{-1}({\bf z}-{\bf z}_0)},
\end{equation}
where ${\bf z}_0$ and $\Gamma$ are variational parameters.  As for the case $M=1$, these 
parameters are obtained by enforcing self-consistency.
To obtain relevant identities, we invoke two alternative formulations of the partition function:
\ba
\Xi &=& 
\sum_{\{N_i\}}\prod_{i=1}^{M}\bigg(\frac{\lambda_{i}^{N_i}}{N_i!}\bigg)\,
e^{-\frac{1}{2}({\bf N}-\boldsymbol{\varrho})^{T}{\mathcal E}({\bf N}-\boldsymbol{\varrho})}\nonumber\\
&=&\int \frac{d{\bf z}}{\sqrt{\det{2\pi\mathcal E}}}\,e^{-S({\bf z})-i\boldsymbol{\varrho}\cdot{\bf z}}.  
\label{eq:Xi_N}
\ea
An analogous relation to that in Eq. (\ref{eq:identity_1}) is
\be
\lambda_i\frac{\partial\log\Xi}{\partial\lambda_i} = 
\sum_{j=1}^M{\mathcal E}^{-1}_{ij}\frac{\partial\log\Xi}{\partial\varrho_j} + \varrho_i.
\label{eq:ident_M}
\ee
Other identities follow by repeated application of 
$\frac{\partial }{\partial \varrho_i}$ or $\frac{\partial }{\partial \lambda_i}$.  The first three identities are:
 \ba
 &&\lambda_i\big\langle e^{iz_i}\big\rangle = -\sum_{j=1}^M{\mathcal E}^{-1}_{ij}\langle iz_j\rangle\nonumber\\
 &&-\lambda_i\big\langle i\delta z_j \delta e^{iz_i}\big\rangle = \delta_{ij} 
 - \sum_{k=1}^M{\mathcal E}^{-1}_{ik}\langle \delta z_k\delta z_j\rangle
 \nonumber\\
 &&\big\langle e^{iz_i}\big\rangle\delta_{ij} + \lambda_i\big\langle \delta e^{iz_i}\delta e^{iz_j}\big\rangle 
= -\sum_{k=1}^M{\mathcal E}^{-1}_{ik}\langle i\delta z_k \delta e^{iz_j}\rangle
 \nonumber\\&&~~~~~~~~~~~~~~~~~~~~~~~\dots
\label{eq:relations_M}
\ea   
and the relevant expectation values for a Gaussian system are: 
\ba
&&\langle iz_i\rangle_g = -y_{i}\nonumber\\
&&\big\langle e^{iz_i}\big\rangle_g = e^{-y_{i}}e^{-\Gamma_{ii}/2}\nonumber\\
&&\langle \delta z_i\delta z_j\rangle_g = \Gamma_{ij}\nonumber\\
&&\big\langle i\delta z_j\delta e^{iz_i}\big\rangle_g = -e^{-y_{i}}e^{-\Gamma_{ii}/2}{\Gamma}_{ij}\nonumber\\
&&\big\langle \delta e^{iz_i} \delta e^{iz_j}\big\rangle_g = \big(e^{-y_{i}}e^{-\Gamma_{ii}/2}\big)\big(e^{-y_{j}}
e^{-\Gamma_{jj}/2}\big)\big(e^{-\Gamma_{ij}}-1\big).
\nonumber\\
\label{eq:values_gauss}
\ea
Equation two is a linear version of equation three.  
Substituting these into equations in Eq. (\ref{eq:relations_M}) we get
\ba
&&\sum_{j=1}^M{\mathcal E}^{-1}_{ij} y_{j} = \lambda_ie^{-y_{i}}e^{-\Gamma_{ii}/2}\nonumber\\
&&\sum_{k=1}^M{\mathcal E}^{-1}_{ik}\Gamma_{kj} + \lambda_i e^{-y_{i}}e^{-\Gamma_{ii}/2}\Gamma_{ij} = \delta_{ij} \nonumber\\
&&\sum_{k=1}^M{\mathcal E}^{-1}_{ik}\Gamma_{kj} +\lambda_{i}e^{-y_{i}}e^{-\Gamma_{ii}/2}\big(1-e^{-\Gamma_{ij}}\big)=\delta_{ij}
 \nonumber\\&&~~~~~~~~~~~~~~~~~~~~~~~\dots
\label{eq:var_M}
\nonumber\\
\ea

To retain simplicity, we limit our analysis to a two component system with the following simple interaction 
matrix:
\be
{\mathcal E}=\varepsilon
\begin{bmatrix*}[r]
1  & -1 \\
-1 & 1  
\end{bmatrix*}.
\label{eq:EPN2}
\ee
(Note that the matrix ${\mathcal E}$ is singular, but this does not pose problems for obtaining 
physically meaningful results as ${\mathcal E}^{-1}\Gamma$ is no longer singular.)
Particles of the same species repel and particles of different species attract each other.  The average 
total number of particles is fixed, $\langle (N_1+N_2)\rangle=N_T$.  Furthermore, we set 
$\lambda_i=\lambda$. With these constraints $y_i=0$, and it only remains to obtain the matrix $\Gamma$.
The variational construction based on equations $(1,2)$ in Eq. (\ref{eq:var_M}) obtains an analytical solution,
\be
{\Gamma}=\frac{\varepsilon}{1+2\varepsilon N}
\begin{bmatrix*}[r]
1  & -1 \\
-1 & 1  
\end{bmatrix*},
\label{eq:Gamma}
\ee
related to fluctuations in a particle number via 
\be
\big\langle\delta N_i\delta N_j\big\rangle = N\delta_{ij} - N^2\Gamma_{ij}.
\ee
The variational construction based on equations $(1,3)$ does not admit analytical solution and 
data points are obtained numerically.

\begin{figure}[h] 
 \begin{center}%
 \begin{tabular}{rr}
  \includegraphics[height=0.25\textwidth,width=0.4\textwidth]{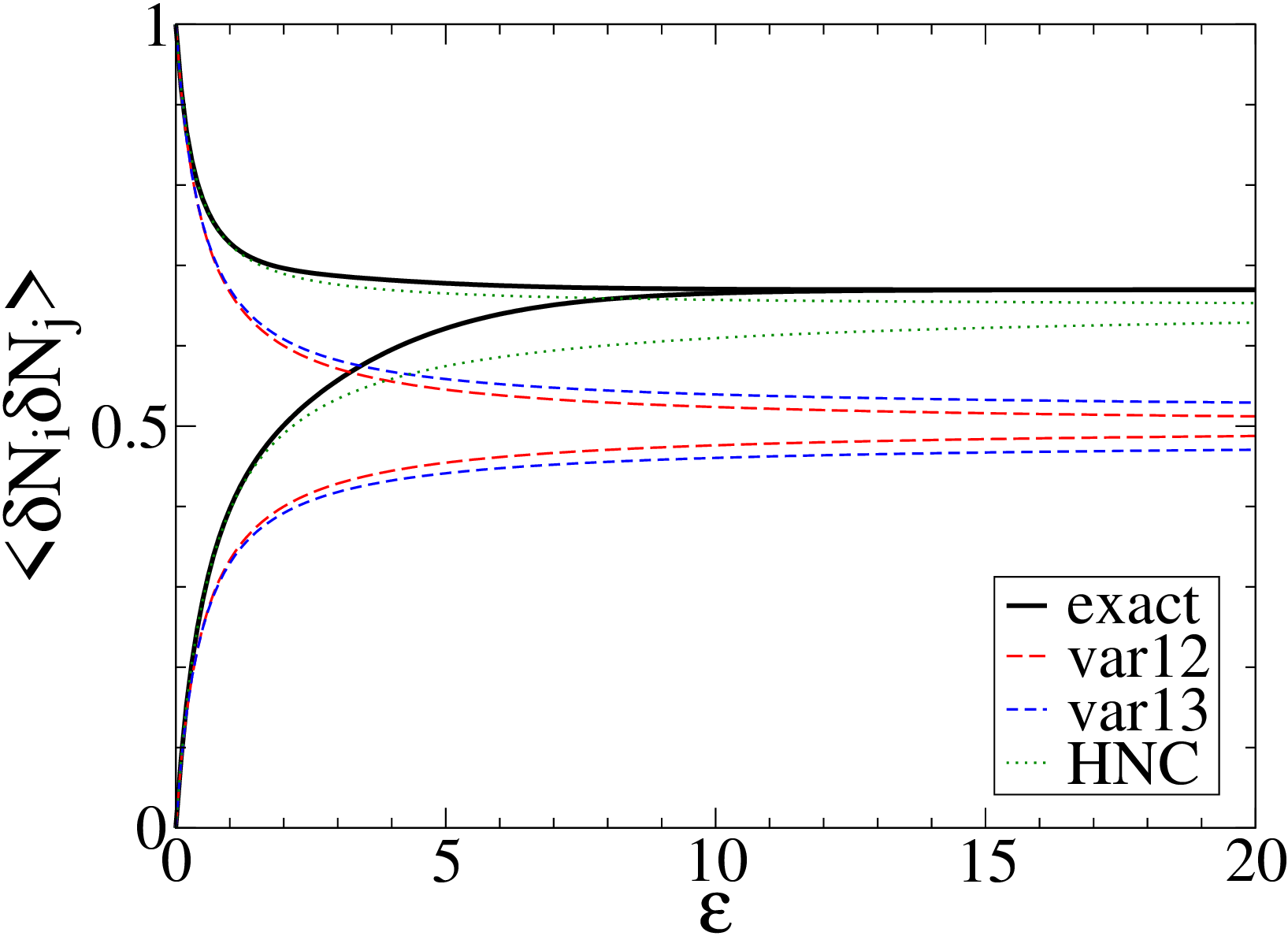}\\
\end{tabular}
 \end{center}
\caption{The particle number fluctuations as a function of $\varepsilon$.  The total
number of particles is kept fixed at $N_T=2$.  The upper curves represent the same-species
and the lower curve the cross-species fluctuations.  }
\label{fig:NiNj}
\end{figure}

Fig. (\ref{fig:NiNj}) plots different results for $\langle\delta N_i\delta N_j\rangle$.
The upper curves represent correlation between the same and the lower
curves between opposite species.  Their conjoining at large $\varepsilon$ implies that
different species form permanent pairs.  The onset of pair formation is captured 
by both self-consistent schemes, but permanent pairs form only in the limit 
$\varepsilon\to\infty$, while the exact results indicate that pairing sets in much faster and is
completed around 
$\varepsilon=10$.  The results for different variational schemes are not drastically 
different, but the gap between curves is smaller for the scheme $(1,2)$.
\graphicspath{{figures/}}

\subsection{digression to the liquid-state theory}
Given the present simple model, we make digression into the liquid-state theories based on the 
Ornstein-Zernike equation.  The relevant quantity 
in the liquid-state formalism is the correlation function, $h_{ij}$, which for the present system is
\begin{equation}
h_{ij} = \frac{\langle \delta N_i \delta N_j\rangle}{\langle N_i\rangle\langle N_j\rangle}
-\frac{\delta_{ij}}{\langle N_j\rangle}.
\label{eq:h_ij}
\end{equation}
The Kronecker delta function subtracts interactions of a particle with itself included in the first 
term.  The Ornstein-Zernike equation relates $h_{ij}$ to the direct 
correlation function $c_{ij}$,
\begin{equation}
h_{ij} = c_{ij} + \sum_{k=1}^M\langle N_k\rangle h_{jk}c_{ki}.
\label{eq:OZ}
\end{equation}
To obtain $c_{ij}$, some sort of closure is required, and which, generally, is obtained based on 
another exact relation, 
\begin{equation}
h_{ij} = e^{-{\mathcal E}_{ij}+h_{ij}-c_{ij} + {\mathcal B}_{ij}} - 1,
\label{eq:B}
\end{equation}
which is a standard result in the integral equation theories and which introduces the bridge function 
${\mathcal B}_{ij}$ \cite{Hansen}.  Within the hypernetted chain approximation 
(HNC) ${\mathcal B}_{ij}=0$.  For inhomogeneous systems an additional closure 
for $\langle N_i\rangle$ is needed, but as the present system is homogeneous, 
$\langle N_i\rangle=N_T/2$, there is no need of a third relation.  The two coupled equations to be solved
are:
\ba
&&h_{ij} = c_{ij} + \frac{N_T}{M}\sum_{k=1}^M h_{jk}c_{ki},\nonumber\\
&&h_{ij} = e^{-{\mathcal E}_{ij}+h_{ij}-c_{ij}} - 1,\nonumber\\
\ea
and the results are shown in Fig. (\ref{fig:NiNj}).  Although 
the gap between different curves closes only in the limit $\varepsilon\to \infty$, precluding 
formation of permanent pairs, the absolute asymptotic value is closer to the exact curve.  

\section{One-component plasma}
\label{sec4}

In this final section we repeat the steps developed for the toy model on a Coulomb system. 
We begin with a general partition function,
\be
\Xi = \sum_{N=0}^{\infty}\frac{\lambda^N}{N!}\int d{\bf r}_1\dots\int d{\bf r}_N\, e^{-{\mathcal H}_N},
\ee
where the dimensionless Hamiltonian is 
\be
{\mathcal H}_N = \frac{1}{2}\int d{\bf r}\int d{\bf r}'\,\hat\rho({\bf r}){\mathcal E}({\bf r},{\bf r}')\hat\rho({\bf r}') 
+ \int d{\bf r}\,\hat\rho({\bf r})\beta U({\bf r}),
\label{eq:H}
\ee
\be
\hat\rho({\bf r}) = \sum_{i=1}^N\delta({\bf r}-{\bf r}'),
\ee
is the density operator, $U({\bf r})$ is an external potential, ${\mathcal E}({\bf r},{\bf r}')$ denotes 
inter-particle interactions, which for Coulomb particles with charge $q$ and inside a medium with 
dielectric constant $\epsilon$ is
\be
{\mathcal E}({\bf r},{\bf r}') = \frac{\beta q^2}{4\pi\epsilon |{\bf r}-{\bf r}'|},
\ee
and 
\be
\lambda = \frac{e^{\beta\mu}e^{{\mathcal E}({\bf r},{\bf r})/2}}{\Lambda^3}
\ee
is the normalized fugacity that subtracts self-interactions in Eq. (\ref{eq:H}).

Using a formal identity analogous to that in Eq. (\ref{eq:identity}), 
\ba
&&\int\!\!\frac{{\mathcal D}\phi}{\sqrt{\det 2\pi {\mathcal E}}}
e^{-\frac{1}{2}\int d{\bf r}\int d{\bf r}'\,\phi({\bf r}){\mathcal E}^{-1}({\bf r},{\bf r}')\phi({\bf r}')}
e^{i\int d{\bf r}\hat\rho({\bf r})\phi({\bf r})}\nonumber\\
&&=e^{-\frac{1}{2}\int d{\bf r}\int d{\bf r}'\,\hat\rho({\bf r}){\mathcal E}({\bf r},{\bf r}')\hat\rho({\bf r}')},
\ea
but where the integral is a functional integral over a fluctuating field $\phi({\bf r})$, and 
the determinant is the functional determinant, leads to a field-theoretical formulation of a partition function
\ba
\Xi = \int {\mathcal D}\phi\,e^{-{\mathcal S}[\phi]},
\ea
where the action is
\ba
{\mathcal S}[\phi] &=& 
\frac{1}{2}\int d{\bf r}\int d{\bf r}'\phi({\bf r}){\mathcal E}^{-1}({\bf r},{\bf r}')\phi({\bf r}')\nonumber\\
&-& \int d{\bf r}\,\lambda e^{-\beta U({\bf r})} e^{i\phi({\bf r})}.  
\ea
and 
\be
{\mathcal E}^{-1}({\bf r},{\bf r}') = -\frac{\epsilon}{\beta q^2}\nabla^2\delta({\bf r}-{\bf r}')
\label{eq:E_inv}
\ee
is the inverse of a Coulomb interaction ${\mathcal E}({\bf r},{\bf r}')$.

\subsection{the saddle-point approximation}
As before, the auxiliary field is generalized to a complex function, $\Phi = \phi + i\psi$.  
The saddle-point,
\be
\frac{\delta S[\Phi]}{\delta \Phi({\bf r})}\bigg|_{\Phi=\Phi_0} = 0,
\ee
yields the following equation
\be
\int d{\bf r}'\,{\mathcal E}^{-1}({\bf r},{\bf r}')\Phi_0({\bf r}') = i\lambda e^{-\beta U({\bf r})}e^{i\Phi_0({\bf r})}. 
\ee
where the solution is strictly imaginary, $\Phi_0=i\psi_0$, and by using Eq. (\ref{eq:E_inv}) we get
\be
\nabla^2\psi_0({\bf r}) = -\lambda \bigg(\frac{\beta q^2}{\epsilon}\bigg) e^{-\beta U({\bf r})}e^{-\psi_0({\bf r})},
\ee  
where $\psi_0$ is the reduced electrostatic potential related to a true electrostatic potential $\Psi$ as
$\psi_0=\beta q\Psi$.  The resulting equation is the Poisson-Boltzmann equation for a one-component 
plasma in an external potential $U({\bf r})$.

\subsection{non-perturbative approach}
As for the toy model, we formulate non-perturbative approach through enforcing self-consistency.
We first generate the relevant identities between expectation values.  By introducing a source term,
\be
S[\phi] \to S[\phi] + i\int d{\bf r}\,\varrho({\bf r})\phi({\bf r}), 
\ee
the Hamiltonian becomes  
\ba
{\mathcal H}_N &=& 
\frac{1}{2}\int d{\bf r}\int d{\bf r}'\,
\big[\hat\rho({\bf r})-\varrho({\bf r})\big]C({\bf r},{\bf r}')\big[\hat\rho({\bf r}')-\varrho({\bf r}')\big] \nonumber\\
&+&\int d{\bf r}\,\hat\rho({\bf r})\beta U({\bf r}),
\label{eq:H2}
\ea
Identities analogous to Eq. (\ref{eq:ident_M}) are generated from the relation
\be
\frac{1}{\beta}\frac{\delta\log\Xi}{\delta U({\bf r})} = 
\bigg(\frac{\epsilon}{q^2\beta}\bigg)\nabla^2\bigg(\frac{\delta\log\Xi}{\delta\varrho({\bf r})}\bigg) 
- \varrho({\bf r}).%
\ee
The first three identities, after setting $U({\bf r})$ and $\varrho({\bf r})$ to zero, are:
\ba
&&\nabla^2\Big\langle i\Phi({\bf r})\Big\rangle 
= \lambda\bigg(\frac{\beta q^2}{\epsilon}\bigg)\Big\langle e^{i\Phi({\bf r})}\Big\rangle
\nonumber\\
&&\bigg(\frac{\epsilon}{q^2\beta}\bigg)\nabla^2\Big\langle \delta\Phi({\bf r})\delta\Phi({\bf r}')\Big\rangle 
+\lambda\Big\langle i\delta\Phi({\bf r}')\delta e^{i\Phi({\bf r})}\Big\rangle\nonumber\\ 
&&~~~~~~~~~~~~~~~~~~~~~~~~~~~~~~~~~~~~~~~~~~~~~~~~~
= -\delta({\bf r}-{\bf r}')\nonumber\\
&&\delta({\bf r}-{\bf r}')\Big\langle e^{i\Phi({\bf r})}\Big\rangle
+ \lambda\Big\langle\delta e^{i\Phi({\bf r})}\delta e^{i\Phi({\bf r}')}\Big\rangle
\nonumber\\
&&~~~~~~~~~~~~~~~~~~~~~~~~~~~~
= \bigg(\frac{\epsilon}{q^2\beta}\bigg)\nabla^2
\Big\langle i\delta\Phi({\bf r})\delta e^{i\Phi({\bf r}')}\Big\rangle\nonumber\\
&&~~~~~~~~~~~~~~~~~~~~~~~~~~\dots
\label{eq:identities_field}
\ea
where
\be
i\delta \Phi({\bf r}) = i\Phi({\bf r}) - \big\langle i\Phi({\bf r})\big\rangle,
\ee 
and
\be
\delta e^{i\Phi({\bf r})} = e^{i\Phi({\bf r})} - \big\langle e^{i\Phi({\bf r})}\big\rangle.
\ee
For a Gaussian reference partition function,
\be
\Xi_g = \int {\mathcal D}\Phi\,e^{-\frac{1}{2}(\Phi({\bf r})-\Phi_0({\bf r}))
\Gamma^{-1}({\bf r},{\bf r}')(\Phi({\bf r}')-\Phi_0({\bf r}'))}.
\ee
the relevant expectation values are,
\ba
&&\langle i\Phi({\bf r})\rangle_{g} = -\psi_0({\bf r})\nonumber\\
&&\langle e^{i\Phi({\bf r})}\rangle_g = e^{-\psi_0({\bf r})}e^{-\Gamma({\bf r},{\bf r})/2}  \nonumber\\
&&\langle \delta\Phi({\bf r})\delta\Phi({\bf r}')\rangle_{g} = \Gamma({\bf r},{\bf r}')\nonumber\\
&&\langle i\delta\Phi({\bf r}')\delta e^{i\Phi({\bf r})}\rangle = 
-e^{-\psi_0({\bf r})}e^{-\Gamma({\bf r},{\bf r})/2}\Gamma({\bf r},{\bf r}')
\nonumber\\
&&\langle \delta e^{i\Phi({\bf r}})\delta e^{i\Phi({\bf r}')}\rangle 
= \big(e^{-\psi_0({\bf r})}e^{-\Gamma({\bf r},{\bf r})/2}\big)\nonumber\\
&&~~~~~~~~~~~~~~~~~~~~\times
\big(e^{-\psi_0({\bf r}')}e^{-\Gamma({\bf r}',{\bf r}')/2}\big)\big(e^{-\Gamma({\bf r},{\bf r}')}-1\big).\nonumber\\
\ea
The three relations in Eq. (\ref{eq:identities_field}) become
\ba
&&\nabla^2\psi_0({\bf r}) = 
-\lambda\bigg(\frac{\beta q^2}{\epsilon}\bigg)e^{-\psi_0({\bf r})}e^{-\Gamma({\bf r},{\bf r})/2}\nonumber\\
&&\nabla^2\Gamma({\bf r},{\bf r}') 
= \lambda\bigg(\frac{\beta q^2}{\epsilon}\bigg)
e^{-\psi_0({\bf r})}e^{-\Gamma({\bf r},{\bf r})/2}\Gamma({\bf r},{\bf r}')\nonumber\\
&&~~~~~~~~~~~~~-
\bigg(\frac{\beta q^2}{\epsilon}\bigg)\delta({\bf r}-{\bf r}')\nonumber\\
&&\nabla^2\Gamma({\bf r},{\bf r}') 
= \lambda\bigg(\frac{\beta q^2}{\epsilon}\bigg)e^{-\psi_0({\bf r})}
e^{-\Gamma({\bf r},{\bf r})/2}\bigg(1-e^{-\Gamma({\bf r},{\bf r}')}\bigg)
\nonumber\\
&&~~~~~~~~~~~~~- \bigg(\frac{\beta q^2}{\epsilon}\bigg)\delta({\bf r}-{\bf r}').\nonumber\\
\label{eq:charges}
\ea

\section{Conclusion}
Using a very simple model we review the basic steps for deriving the field-theoretical formulation
of statistical mechanics.  To avoid using the GBF inequality (inapplicable for complex actions) in constructing 
a non-perturbative approach, we explore alternative schemes.  Within the scheme proposed 
in this work, self-consistency is enforced by explicit satisfaction of a number of exact identities, to obtain
which we provide a recipe.  The GBF variational construction is found to satisfy the first two relations 
in the hierarchy.  This provides the GBF variational scheme with somewhat different
interpretation.  One can also chose different identities in the hierarchy, leading to different type
of self-consistency and approximation.  

\begin{acknowledgments}
This work was supported by the agence nationale de la recherche via the project FSCF.  
\end{acknowledgments}



\end{document}